\documentclass[]{spie}  

\usepackage{amsmath,amsfonts,amssymb}
\usepackage{graphicx}
\usepackage{float}
\usepackage[colorlinks=true, allcolors=blue]{hyperref}

\title{Tight focal spots using azimuthally polarised light from a Fresnel cone}

\author[a]{R. D. Hawley}
\author[a]{R. Offer}
\author[a]{N. Radwell}
\author[a]{S. Franke-Arnold}
\affil[a]{SUPA and School of Physics and Astronomy, University of Glasgow, Kelvin Building, Renfrewshire, Glasgow, G12 8QQ, UK}

\authorinfo{Further author information: (Send correspondence to R. D. Hawley)\\E-mail: r.hawley.1@research.gla.ac.uk}

\begin{document} 
\maketitle

\begin{abstract}
When focusing a light beam at high numerical aperture, the resulting electric field profile in the focal plane depends on the transverse polarisation profile, as interference between different parts of the beam needs to be taken into account.  It is well known that radial polarised light produces a longitudinal polarisation component and can be focused below the conventional diffraction limit for homogeneously polarised light, and azimuthally polarised light that carries one unit of angular momentum can achieve even tighter focal spots. This is of interest for example for enhancing resolution in scanning microscopy. 
There are numerous ways to generate such polarisation structures, however, setups can be expensive and usually rely on birefringent components, hence prohibiting broadband operation. We have recently demonstrated a passive, low-cost technique using a simple glass cone (Fresnel cone) to generate beams with structured polarisation. We show here that the polarisation structure generated by Fresnel cones focuses better than radial polarised light at all numerical apertures. 
 Furthermore, we investigate in detail the application of polarised light structures for two-photon microscopy.  Specifically we demonstrate a method that allows us to generate the desired polarisation structure at the back aperture of the microscope by pre-compensating any detrimental phase shifts using a combination of waveplates. 
\end{abstract}


\section{INTRODUCTION}
\label{sec:intro}  

By convention the polarisation of light refers to the direction of oscillation of its electric field component. Our eyes are generally not able to detect polarisation\footnote{With the exception of the ``Haidinger's brush'' phenomenon.} and conventional cameras are insensitive to it, however, its propagation and interactions are applied in many areas such as in optical communication~\cite{Hayee2001,Wang2014}, optical data storage~\cite{Barrera2006,Li2011}, remote sensing~\cite{Tyo2006c,Diner2007} and in biomedical research~\cite{Ushenko2008,Ghosh2011,Alali2015}. Many optical devices in these fields use lenses to focus light, where usually the polarisation of light does not need to be considered, however, fascinating effects are found when focusing in a high numerical aperture (NA) system. For example (and perhaps surprisingly), a linearly polarised beam with a Gaussian intensity profile focuses to an elliptical spot shape~\cite{Dorn2003c}. Circularly polarised light can be used to achieve a circular spot, and this spot size is often associated with the conventional diffraction limit. The strong focusing of beams of light with more interesting polarisation structures have been investigated, where it was shown that a radially polarised beam focuses below the conventional diffraction limit in a high-NA system~\cite{Youngworth2000b,Dorn2003a}. It was more recently shown that an azimuthally polarised beam with orbital angular momentum (OAM) can focus to an even smaller spot than both the conventional-circular and radially polarised beams, and furthermore does not rely on an especially high-NA objective lens to achieve this~\cite{Hao2010}.

Previously, we demonstrated a low-cost method to generate achromatic vector vortex beams~\cite{Radwell2016}, including the azimuthally polarised beam discussed by Hao \textit{et al.}~\cite{Hao2010} that has a $0$-$2\pi$ spiral phase. This was achieved by using the simple back-reflection from a solid glass cone - so-called Fresnel cone. Furthermore, Fresnel cones allow broadband operation, which is ideal for microscopy applications employing tunable lasers (such as in many multi-photon systems) or ultrashort pulses. Fresnel cones have not only been shown to generate polarisation structures, but also to use polarisation structuring to measure the polarisation state of light in single-shot broadband polarimetry~\cite{Hawley2019}.

When attempting to apply polarisation effects in microscopy, an issue arises in that each transmission and reflection of the light beam through the microscope system can cause detrimental polarisation shifts to the initial polarisation state. These shifts vary in significance for different optical components, but perhaps the largest polarisation shifts can be observed when transmitting through dichroic beamsplitters (often found in fluorescent microscope systems). Although these unwanted polarisation shifts are usually due to reflection or transmission through a serious of components, the entire system can be summarised in a single Mueller matrix describing the overall effect on the polarisation of the initial beam. This is useful as a Mueller matrix can be experimentally measured for the overall system, and following this, precompensation of the generated polarisation structure using a half- and quarter-wave plate can be applied (similarly to what was achieved by Chou \textit{et al.}~\cite{Chou2008} for uniformly polarised beams). The result is that these unwanted polarisation shifts caused by the microscope system now essentially correct the precompensated beam, so that the required state arrives at the back aperture of the microscope's objective lens. We apply this precompensation method to an azimuthally polarisation beam which is transmitted through a microscope system, to achieve the desired state at the back aperture plane of the microscope objective.

\section{Fresnel cone polarisation structuring}

A Fresnel cone is a solid glass cone that has a $90^\circ$ apex angle, and was previously investigated as a method for generating beams with structured polarisation and orbital angular momentum (OAM). For light incident at the normal to the front surface of the Fresnel cone, the angle of incidence at the back conical surface is above the critical angle for total internal reflection (TIR) and so all of the light is retro-reflected. When this occurs at a boundary such as this, between glass and air, Fresnel's equations predict a phase shift between the components of the incident polarisation that are perpendicular ($s$) and parallel ($p$) to the plane of incidence at the cone back surface. In addition to the angle of incidence, the amount of phase shift also depends on the refractive index of the cone material, and in our case we use Fresnel cones with a refractive index of $1.556$ (using H-BAK6 glass, developed in combination with Gooch and Housego) to provide a total phase shift of $\sim\frac{\pi}{2}$ for the two combined reflections when using at 850nm (a typical wavelength used for excitation in multi-photon microscopy). For a uniformly polarised incident beam, the $s$ and $p$ decomposition at the back conical surface varies azimuthally around the cone apex. The result is that when considering polarisation, Fresnel cones can be thought of as azimuthally varying wave-plates, with our chosen refractive index corresponding to a quarter-wave plate.

\begin{figure}[h]
 \centering
 \includegraphics[scale=1]{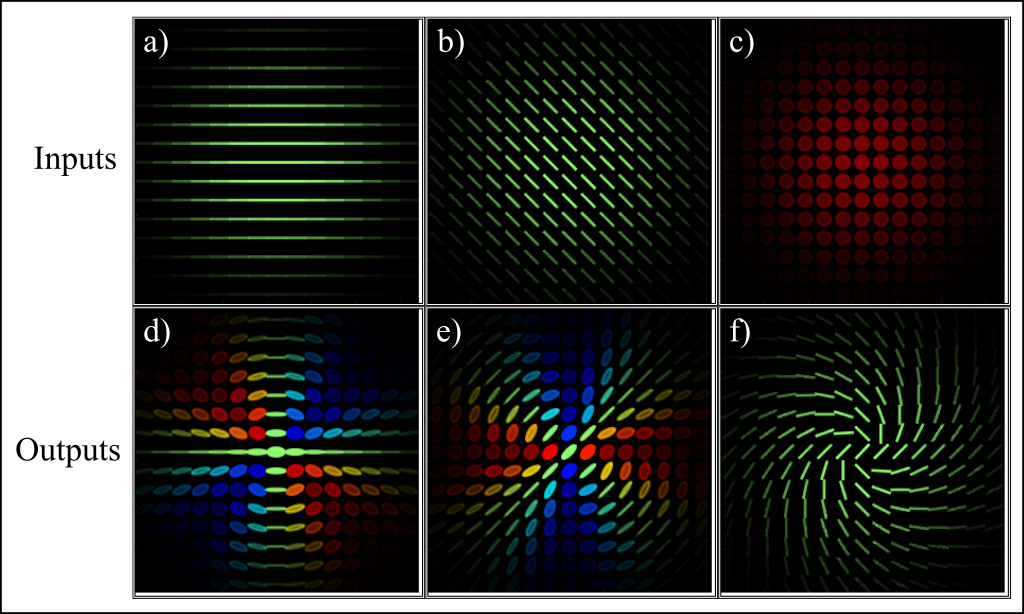}
 \caption{\label{fig:states}Simulated examples for different input polarisation states a)-c) to the Fresnel cone and their associated output states d)-f). Green lines represent linearly polarised light and the degree of red and blue represents right and left handedness, with orientation of the ellipse denoting the orientation of the polarisation.}
\end{figure}

\begin{figure}[h]
 \centering
 \includegraphics[scale=1]{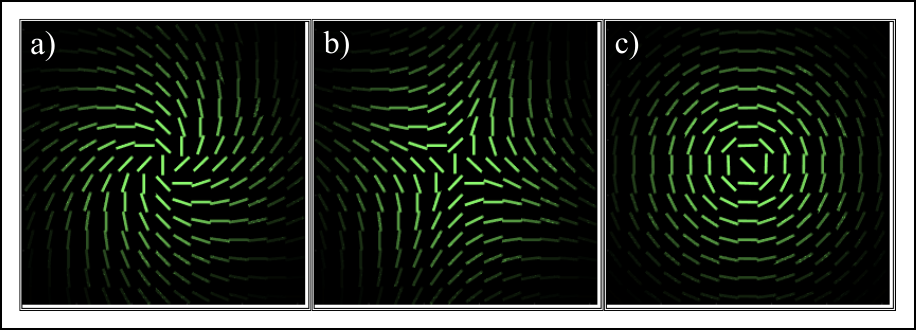}
 \caption{\label{fig:cones}Simulated output polarisation states for a right-handed circular input to the Fresnel cone after different components. a) polarisation state directly after Fresnel cone, b) after beamsplitter reflection and c) after half-wave plate at $22.5^\circ$.}
\end{figure}

The polarisation properties of a Fresnel cone can be calculated using Mueller calculus, through which~$4~\times~1$ Stokes vectors representing the light's input polarisation state are acted upon by~$4~\times~4$ Mueller matrices representing optical components, to determine output states. These matrices are well known for common optical components such as polarisers and wave plates, but to obtain a Mueller matrix for a Fresnel cone the method described by Radwell \textit{et al.}~\cite{Radwell2016} is used, which we repeat here for completeness. First a Jones matrix is constructed using the Jones matrix for total internal reflection in a solid glass wedge as follows,
\begin{equation}
\textbf{J}_{\rm wedge}=
\begin{bmatrix}
 {r_{p}}^{2}& &0\\ 
 0& &{r_{s}}^{2} 
\end{bmatrix}
={r_{p}}^{2}
\begin{bmatrix}
 1& &0\\ 
 0& &e^{i2\delta},
\end{bmatrix},
\label{wedge}
\end{equation}
where
\begin{equation}
\delta=\textrm{arg}(r_s)-\textrm{arg}(r_p)=\textrm{arg} \bigg( \frac{n^{2}+i\sqrt{1-2n^{2}}}{1-n^{2}} \bigg),
\end{equation}
$r_{s}$ and $r_{p}$ are the Fresnel reflection coefficients for $s$ and $p$ polarisation components and $n=n_{\textrm{air}}/n_{\textrm{glass}}$. This is then combined with rotation matrices to form the solid glass cone,
\begin{equation}
\textbf{J}_{\rm cone}=\textbf{R}(-\theta)\textbf{J}_{\rm wedge}\textbf{R}(\theta),
\label{Cone}
\end{equation}
where $\textbf{R}(\theta)$ is the Jones matrix for a rotation by $\theta$, the azimuthal angle around the cone apex from the horizontal, given as
\begin{equation}
\textbf{R}(\theta)=
\begin{bmatrix}
\cos{\theta}& &\sin{\theta}\\ 
-\sin{\theta}& &\cos{\theta} 
\end{bmatrix}.
\label{rotation}
\end{equation}
The Mueller matrix can then be determined by following the method described by Azzam and Bashara~\cite{Azzam1987}. We then find the theoretical Mueller matrix for a Fresnel cone with a total phase shift of $\frac{\pi}{2}$ to be
\begin{equation}
\textbf{M}_{\rm cone}=
\begin{bmatrix}
1& &0& &0& &0\\ 
0& &\cos^2{2\theta}& &\frac{1}{2}\sin{4\theta}& &-\sin{2\theta}\\
0& &\frac{1}{2}\sin{4\theta}& &\sin^2{2\theta}& &\cos{2\theta}\\
0& &\sin{2\theta}& &-\cos{2\theta}& &0
\end{bmatrix}.
\label{rotation_exp}
\end{equation}
 Using Jones and Mueller calculus, the spatially dependent output polarisation state from a Fresnel cone can be calculated and examples of the results for different inputs are shown in~\autoref{fig:states}.
Experimentally, a non-polarising beamsplitter is the most straight-forward way to couple the light in and out of the Fresnel cone (as shown in~\autoref{fig:setup}). This means that the actual measurable polarisation state will also include a reflection from the beam-splitter. In order to generate radial or azimuthally polarised light, a half-wave plate is used at either $+22.5^\circ$ or $-22.5^\circ$ (see~\autoref{fig:cones} for the polarisation states for each component after the Fresnel cone).
The output state can easily be switched from azimuthal polarisation shown in~\autoref{fig:cones}c) to radial polarisation by either changing the input state from right-handed to left-handed circular, or by rotating the final half-wave plate from $+22.5^\circ$ to $-22.5^\circ$.

\begin{figure}[h]
 \centering
 \includegraphics[scale=0.8]{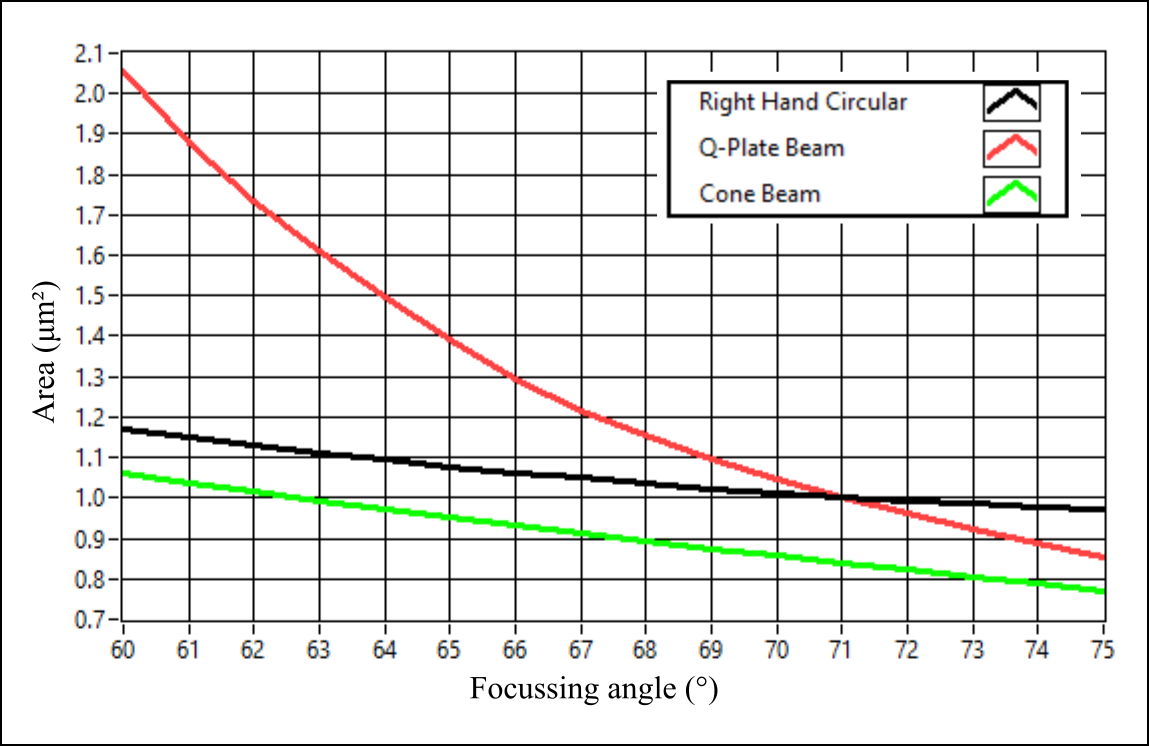}
 \caption{\label{fig:comparison}A comparison of the transverse focal area for a diffraction-limited circularly polarised beam (black), conventional radially polarised beam (red) and azimuthally polarised vector vortex beam from a Fresnel cone (green).}
\end{figure}

The strong focusing properties of these vector vortex beams is simulated, based on vector diffraction theory developed by Richards and Wolf~\cite{Richards1959}. A comparison of the transverse focal area for a diffraction-limited right-handed circularly polarised beam, conventional radially polarised beam and azimuthally polarised beam generated using a Fresnel cone is shown in~\autoref{fig:comparison}. These results agree with previous literature, showing that for a strong enough focusing angle (high enough NA) a radially polarised beam focuses smaller than the diffraction-limited circularly polarised beam, and an azimuthally polarised beam with OAM focuses to a smaller spot than both of these (without the high-NA requirement), here shown for simulations based on light at $1\mu m$ wavelength.
\section{Mueller matrix compensation}
\label{Mueller}

\begin{figure}[H]
 \centering
 \includegraphics[scale=1]{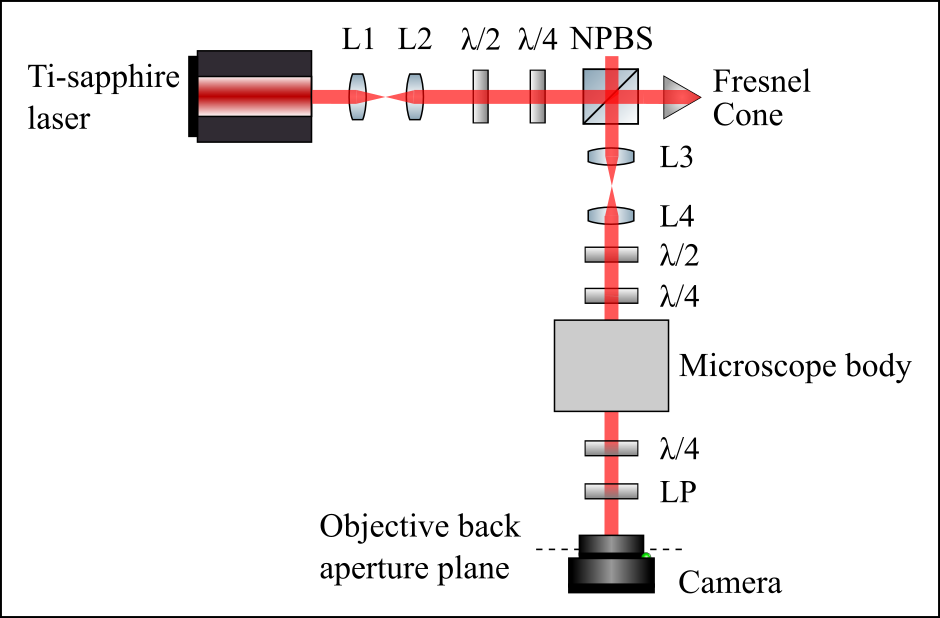}
 \caption{\label{fig:setup}Experimental setup used to generate, compensate and measure structured polarisation beams generated using a Fresnel cone, through a multi-photon microscope system. L1, L2, L3, L4 are lenses, $\lambda /2$ and $\lambda /4$ are half- and quarter-wave plates respectively (for use at $850 nm$). LP is a linear polariser and NPBS is a non-polarising beamsplitter. The microscope body consists of numerous mirrors, lenses and dichroic beamsplitters.}
\end{figure}

The experimental setup for generating, compensating and measuring structured polarisation beams from a Fresnel cone is shown in~\autoref{fig:setup}. These beams are expected to be beneficial in many optical microscope systems, however, the aim of this project is to eventually use these beams in multi-photon fluorescence microscopy, where the benefit of broadband operation of a Fresnel cone can be demonstrated. A Ti-sapphire laser (Chameleon Vision, Coherent), tuned to a wavelength of $850 nm$ (a typical wavelength used in two-photon microscopy), is expanded using lenses L1 and L2 to fill the Fresnel cone front surface. A half- ($\lambda /2$) and quarter-wave ($\lambda /4$) plate are used to generate a circular input to the Fresnel cone (as well as other input states for analysis). A non-polarising beamsplitter (BS014, Thorlabs) is used to couple the light in and out of the Fresnel cone. Two lenses, L3 and L4 then image the back surface of the cone to the galvanometer mirror system, which is imaged to the back aperture of the objective lens through the microscope body. Inside the microscope body there are a number of mirrors, lenses and dichroic beamsplitters, causing unwanted phase shifts to the input polarisation state.

In order to characterise the polarisation properties of the microscope system, the Mueller matrix was measured for the system as a whole. This was done using the method outlined by Fujiwara~\cite{Fujiwara2007}. For a perfect theoretical transmission, the Mueller matrix of the microscope system would ideally be the identity matrix, however, the measured Mueller matrix (normalised) is
\begin{equation}
\begin{bmatrix}
 1& &0& &0& &0\\ 
 0.00260793& &0.92820968& &0.155126027& &0.1647746\\ 
 0.023598868& &-0.062604573& &-0.846448676& &0.347155164\\ 
 0.032967732& &0.08721139& &0.306293706& &0.733688886\\ 
\end{bmatrix}.
\label{2p}
\end{equation}
It is clear from this measurement that the polarisation transmitted through the microscope system is detrimentally shifted from the initial input state. In order to compensate for this the method outlined by Chou \textit{et al.}~\cite{Chou2008} is followed, by using a half- and quarter-wave plate after Fresnel cone beam generation. By measuring the actual structured polarisation state at the back aperture plane of the microscope objective, the required waveplate angles for precompensation can be determined. This is done by calculating the resulting structured polarisation state after rotating compensation wave plates at all angles, and using the settings that produce the polarisation result with the least squares error from the desired azimuthally polarised state.
\begin{figure}[H]
 \centering
 \includegraphics[scale=1]{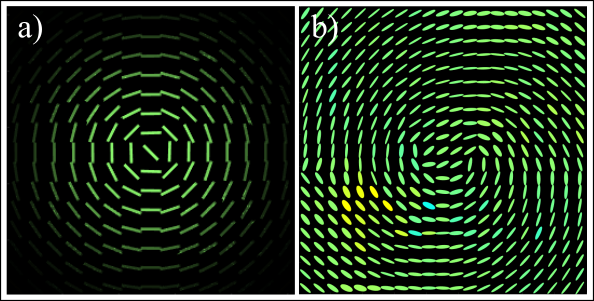}
 \caption{\label{fig:result}a) Ideal polarisation state from theory and b) initial result of azimuthally polarised cone beam at the back aperture plane of the microscope objective after precompensation.}
\end{figure}
Polarisation measurements are made at the back aperture of the microscope objective by using a quarter-wave plate and linear polariser to perform Stokes measurements - projecting the beam into six polarisation basis states, horizontal linear, vertical linear, diagonal linear, anti-diagonal linear and right and left circular, and recording intensity images on a camera. Initial results show that an approximate azimuthally polarised state can be recovered at the back aperture plane of the microscope objective, however, improvements to the characterisation of the system and more precise calibration of the polarisation components could provide more accurate results.

\section{Conclusion}

In summary, we have confirmed through simulation that a Fresnel cone can be used to generate beams that can focus below the conventional diffraction limit. Unlike the well-studied radially polarised light, these azimuthally polarised beams will work even at moderate NAs. We have further shown that while transmission through microscope systems adds detrimental phase shifts to input polarisation state, this can be pre-compensated by using a correctly aligned half- and quarter-wave plate to correct the structured polarisation state, expanding on the method demonstrated by Chou~\textit{et al.}~\cite{Chou2008} for uniformly polarised beams. Initial experimental results confirm that azimuthally polarised light can be recovered at the back aperture plane of the microscope objective, where future work will be carried out to measure sub-diffraction limited focal spots of these focused beams. Fresnel cones are particularly useful for generating these interesting beams as they are low-cost, robust and allow broadband operation unlike many polarisation optics, which is beneficial for fluorescence microscope systems employing tunable lasers.

\section*{Acknowledgements}
We are grateful for our colleagues Peter MacKay at Gooch and Housego for providing the Fresnel cone, and to Ewan McGhee and Leo Carlin for useful discussions on two-photon microscopy. This work has been supported by EPSRC Quantum Technology Program grant number EP/M01326X/1. R. D. Hawley's work was supported by the EPSRC CDT in Intelligent Sensing and Measurement, Grant Number EP/L016753/1.

\bibliography{SPIE} 
\bibliographystyle{spiebib} 

\end{document}